\documentclass{iau} 

\usepackage{graphicx}
\usepackage{xcolor}
\usepackage[hyphens]{url}
\usepackage{hyperref}
\hypersetup{
     colorlinks  = true,
     linkcolor   = blue, 
     anchorcolor = BrickRed,
     citecolor   = blue,
     filecolor   = blue,
     menucolor   = blue,
     runcolor    = cyan,
     urlcolor    = blue
}
\usepackage[authoryear]{natbib}

\pubyear{2019}
\volume{355} 
\jname{The Realm of the Low-Surface-Brightness Universe}
\editors{D. Valls-Gabaud, I. Trujillo \& S. Okamoto, eds.}

\def\etal{\textit{et al.}}

\title[Ultra-deep imaging amateur telescopes] 
{Ultra-deep imaging with amateur telescopes}

\author[David Mart\'\i nez-Delgado]   
{David Mart\'\i nez-Delgado$^1$
}

\affiliation{$^1$Astronomisches Rechen-Institut, University of Heidelberg, \\ M{\"o}nchhofst. 12-14,
DE-69120, Heidelberg, Germany \\ email: {\tt delgado@ari.uni-heidelberg.de} \\[\affilskip]
}

\begin{document}

\maketitle

\begin{abstract}

Amateur small equipment has demonstrated to be competitive tools to obtain ultra-deep imaging of the outskirts of nearby massive galaxies and to survey vast areas of the sky with unprecedented depth.  Over the last decade, amateur data have revealed, in many cases for the first time, an assortment of  large-scale tidal structures around nearby massive galaxies and have detected hitherto unknown low surface brightness systems in the local Universe that were not detected so far by means of resolved stellar populations or H{\sc i} surveys. In the Local Group,  low-resolution deep images of the Magellanic Clouds with telephoto lenses have found some shell-like features, interpreted as imprints of a recent LMC-SMC interaction. I this review, I discuss these highlights and other important results obtained so far in this new type of collaboration between high-class astrophotographers and professional astronomers in the research topic of galaxy formation and evolution. 


\end{abstract}

\firstsection 

\section{Introduction}

Within the hierarchical framework for galaxy formation, the stellar halos of massive galaxies are expected to form and evolve through a succession of mergers with low-mass systems. Numerical cosmological simulations of galaxy assembly in the  $\Lambda$-Cold Dark Matter ($\Lambda$-CDM) paradigm predict that satellite disruption occurs throughout the lifetime of all massive galaxies and, as a consequence,  their stellar halos at the present day should contain a wide variety of diffuse remnants of disrupted dwarf satellites. These cosmological simulations also predict that significant numbers of stellar streams may still be detectable, with sufficiently deep observations, in the outskirts of the majority of nearby massive galaxies (Bullock \& Johnson 2005; Cooper et al. 2010). Thus, the detection and characterization of faint substructures in the stellar halos and a complete census of survivor satellites in nearby massive galaxies is a vital test of the hierarchical nature of galaxy formation that has not yet been fully exploited, mainly because their extremely low surface brightness makes them challenging to observe. 

 In addition, this scenario is challenged to provide a convincing  explanation for the smaller than predicted number of observed satellite galaxies around Local Group spirals and other nearby galaxies (Klypin  et  al.  1999). However, the deep imaging of nearby galaxies has not yet been pushed to the feasible limit with professional facilities. Since the majority of surviving satellites around massive galaxies have surface  brightness lower than $\mu_{g}>25$  mag/arcsec$^{2}$, they remain undetected due to the shallow images available from the Sloan Digital Sky Survey (SDSS) or the  PanSTARRs.  

Over the last decade, small amateur telescopes have demonstrated they can be competitive instruments to undertake ultra-deep imaging studies of the outskirts of nearby galaxies. This offers an alternative and low cost approach for the discovery of diffuse stellar halo substructure predicted by the $\Lambda$-CDM simulations and for completing the dwarf galaxy census around massive galaxies in the local universe.
\begin{figure}[t]
\begin{center}
 \includegraphics[width=1.0\textwidth]{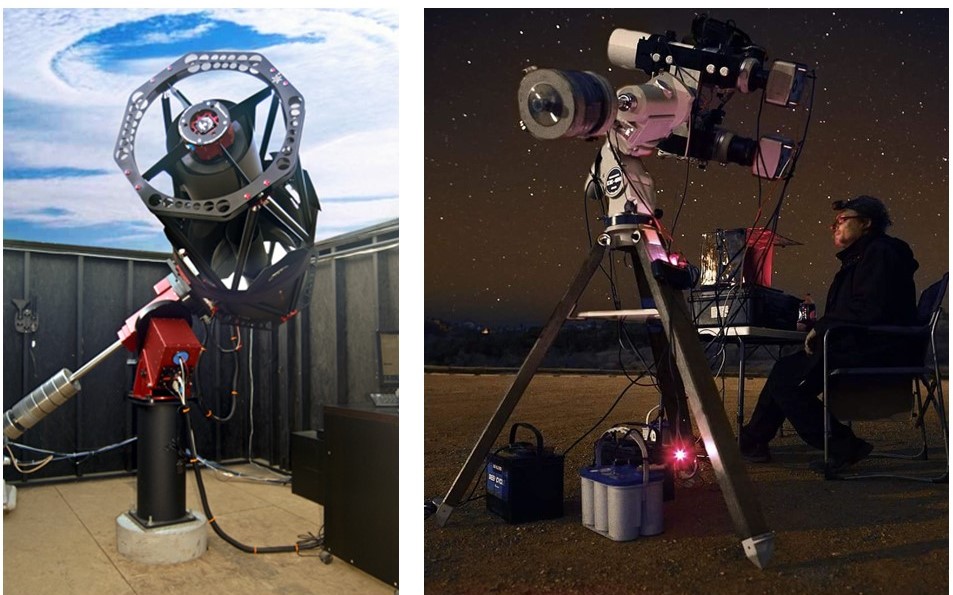} 
 \caption{\footnotesize{Two examples of the amateur equipment used in ultra-deep imaging of nearby galaxies shown in this review: ({\it Left}): The Officina Stellare PRO RC 500 half-meter telescope from the Black Bird II Observatory, located at 1,410 meters elevation in the California Sierra-Nevada Mountains near Alder Springs, California.({\it Right}): Portable equipment composed by a double Takahashi FSQ106EDX, each one with a CCD camera SBIG STK11k CCDs. (Image credits: R.J. Gabany \& R. Bernal)}}
  \label{lugaro:fig1}
\end{center}
\end{figure}




\section{Equipment and Observational Strategy}




The  ultra-deep, wide-field imaging observations of nearby spiral galaxies and selected sky areas described in this review were mainly obtained with privately owned observatories equipped with modest-sized telescopes (0.1-0.8-meter; see Fig. ~1) operating under very dark skies, in collaboration with high-class astrophotographers from Europe, the United States and Chile. Each observing location features spectacularly dark, clear skies with a typical seeing below 1.5". These small telescopes (and, in some cases, telephoto lenses; see Sec. 5) 
were coupled with commercially available CCD cameras equipped with the latest generation of single photographic-film sized imaging chips. This equipment can probe vast sky areas with unprecedented depth, reaching surface brightness levels that are $\sim$ 2–3 magnitudes deeper ($\mu_{r,lim} \sim$ 28 mag arcsec−2) than both the classic photographic plate surveys (e.g., the Palomar Observatory  Sky  Survey)  and  the  available  large-scale digital surveys (e.g.  the SDSS).  Their sensitivity, fast operation and lack of competition for observing time typical of professional observatories have placed these low-cost robotic amateur facilities at the front line of ultra-deep imaging and providing a high impact to the research of low surface brightness galactic structure in nearby galaxies.

\begin{figure}[t]
\begin{center}
 \includegraphics[width=1.0\textwidth]{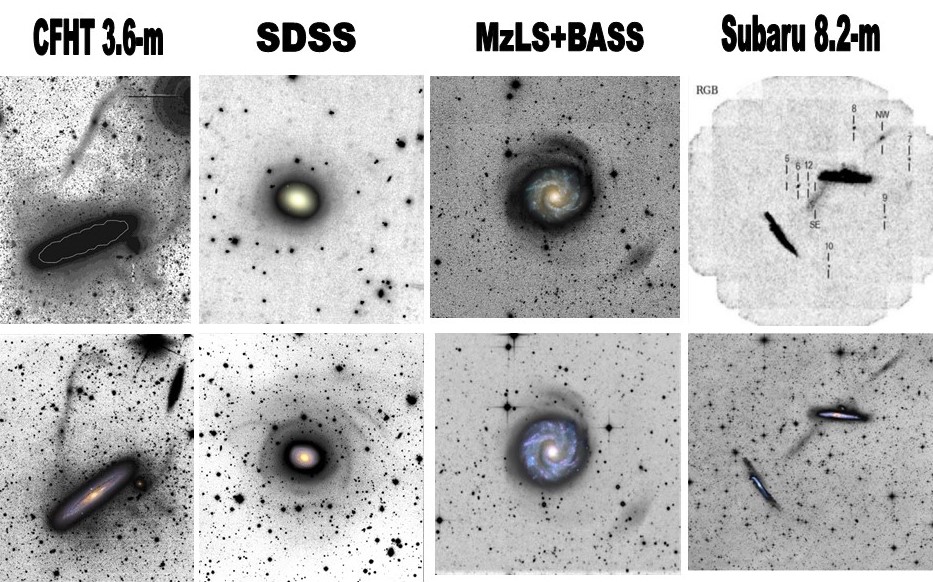} 
 \caption{\footnotesize{A comparison of deep amateur images ({\it bottom panels}) with data obtained with large professional telescopes and large scale CCD imaging surveys ({\it top panels}). From left to right: NGC 4216 (Paudel et al. 2013), NGC 2275 (Morales et al. 2016), NGC 3631 (Mart\'\i nez-Delgado et al. 2019) and NGC 4631 (Tanaka et al. 2017). Amateur images taken by R. Jay Gabany and Karel Teuwen.}}
  \label{lugaro:fig1}
\end{center}
\end{figure}

The observing strategy strives for multiple deep exposures of each target using high throughput clear filters known as luminance (L) filters (4000 \AA $< \lambda <$7000 \AA; see Fig.~1 in Mart\'\i nez-Delgado et al. 2014) and with typical exposure times of  7-8 hours. The typical 3-$\sigma$ SB detection limit (measured in random 2" diameter apertures)  is $\sim$ 28 and 27.5 mag/arcsec$^2$ in $g$ and $r$, that is approximately two magnitudes deeper than the SDSS-II images. The typical 1-$\sigma$ accuracy of the background subtraction (as estimated from the box-to-box variance of the average residual surface brightness) is 29.5 and 28.5 mag/arcsec$^2$ in $g$ and $r$ bands, respectively. From a direct comparison with SDSS data, it was found that these images are ten times deeper than the SDSS ones in terms of photon statistics, with comparable systematic background uncertainties. The magnitudes in the L-filter are transformed to $r$-band with an accuracy of 1\%.

These observations have demonstrated for first time the suitability of small telescopes to detect very faint, diffuse structures in large fields around nearby galaxies. Fig.~2 shows a comparison of some of these amateur telescopes results with some wide-field data produced with professional intermediate-size (3.5--8 meter) telescopes and large scale surveys (SDSS, DECaLs). This illustrates the advantages of small telescopes in detecting features with low surface brightness spread over a large sky area. First,  small short-focal-length telescopes, combined with single chip cameras, covered a larger field of view (30--120 arcmin). The use of these single-chip detectors also made it  easier to flatten the external regions around galaxies in comparison to standard multi-chip detector arrays used with professional telescopes. Furthermore, observations with large telescopes are sometimes subject to significant sky background variations from different sources (e.g. flatfield corrections, photometric zero-points for different chips, fringes, scattered light, reflections, etc). These artifacts complicate the detection of faint structures (see Tal et al. 2009) and their correction adds significant observing time over-head to the data gathering process. Interestingly, Fig.~2 shows the NGC 4631 tidal stream traced by its resolved red giant branch stars obtained with the Subaru 8.2-m telescope (top right panel; Tanaka et al. 2017) is in excellent agreement with the diffuse-light feature visible in a 0.3-meter ROSA telescope image (bottom left panel), confirming that this stream is mainly composed by old-population stars.

\begin{figure}[t]
\begin{center}
 \includegraphics[width=1.0\textwidth]{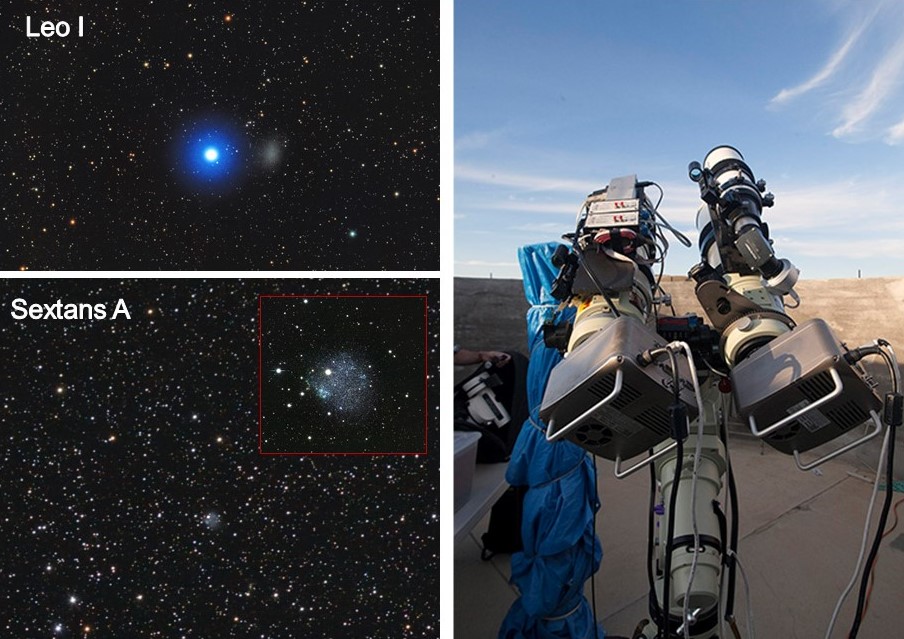} 
 \caption{\footnotesize{{\it Left panel:} Two examples of the results from the observational approach designed to trace diffuse stellar systems as unresolved, rounded over-densities using low-resolution telescopes and telephoto lens (see Sec.~2). In the {\it top panel}, an image of the Leo I dSph taken by Andrea Pistocchini with a Tecnosky Apo triplet 80/480 telescope and a exposure time of  3.8 hours. In the {\it bottom panel}, a cropped image of Sextans dIrr galaxy  taken from a wide-field image obtained by the CEDIC team with a Canon EF 200mm f/2.8L telephoto lens and a total exposure time of 4.5 hours. {\it Right panel:} The double Takahashi FSQ106EDX, each tube with a focal reducer and a CCD camera SBIG STK11k CCDs, is one of the best equipment used in this project to look for unknown dwarf galaxies in the Local Group.}}
  \label{lugaro:fig1}
\end{center}
\end{figure}

Ultra-deep imaging in wide sky areas with portable amateur telescopes and telephoto lenses can also help to complete the census of these hitherto unknown low surface brightness galaxies. In the Local Group spirals, the discoveries of dwarf satellites have been made using stellar density maps of resolved stars, counted in selected areas of the color-magnitude diagrams (CMDs), made from the photometric star catalogs of large scale surveys(e.g. SDSS, Pan-STARRs, DES). In the case of the Milky Way, the main tracers to find these diffuse systems
 are blue stars 2--3 magnitudes fainter than the main sequence (MS)-turnoff of the old stellar population.
However, the relatively shallow photometry of these surveys (with a $g$-band limiting magnitude for point sources of $\sim$ 21.5) and the significant contamination by foreground stars and distant blue galaxies, make it very difficult to complete the census of faint  dwarf companions at distances larger than 100 kpc using this data. For the case of the Andromeda galaxy (and a few Megaparsecs beyond the Local Group), its larger distance makes it prohibitive  to resolve their stellar halo stars that are fainter than the red clump level (e.g. PAndAS photometry; Martin et al. 2009). This means that the M31 satellite population  can only be traced by observations of the less numerous red-giant branch (RGB) stars. This also means dwarf galaxy hunting is certainly biased around the Local Group, where the old populations of low-surface brightness, dwarf spheroidal galaxies can barely be resolved into stars. 

An alternate seach strategy is to observe with instruments that yield stellar systems as difusse unresolved rounded structures (Fig.~3, left panel). Considering that the distance of the most interesting targets (Local Group members) is in the range of 0.1--1 Megaparsec,  it is necessary to use very short-focal ratio instruments (i.e.  f/3 or less) which cannot resolve them into individual stars. For those systems at a distance closer than 100 kiloparsecs (e.g. the Magellanic Clouds, see Sec.~ 4), this low resolution requirement is only obtained with high quality telephoto lenses or apochromatic refractor telescopes equipped with suitable focal reductors (Fig.~3, right panel), that, in the majority of the cases, are installed on portable mounts located in a very dark site (e.g. see left panel in Fig.~1). This search strategy avoids the foreground star and background galaxy contamination that affects the analysis of the stellar density maps extensively used to search for faint satellites in the Milky Way and Andromeda halos. The low-resolution data obtained is also suitable for the same search algorithms and structural or photometric analysis applicable with unresolved galaxies in the Local Universe.

\begin{figure}
\begin{center}
 \includegraphics[width=1.0\textwidth]{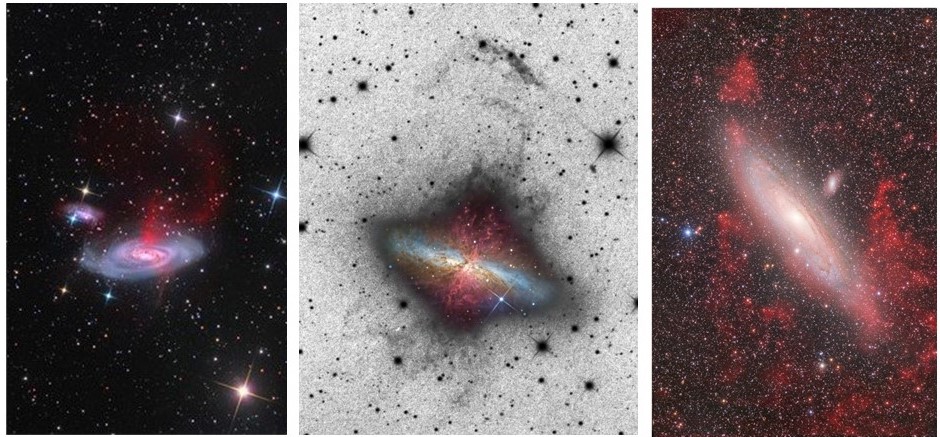} 
 \caption{\footnotesize{Examples of very faint, extended diffuse H$_{\alpha}$ emission (red color) detected around NGC 4569 (left panel), Messier 82 (center panel) and the Andromeda galaxy (right panel) by means of follow-up deep H$_{\alpha}$ observation. The images of NGC 4569 and M82 were taken with a 0.5-m telescope by Mark Hanson and R. Jay GaBany respectively. The M82 is shown in negative with a color inset of the disk of the galaxy. The H$_{\alpha}$ data (red color) from the full color image of the Andromeda galaxy (composed by Giuseppe Donatiello) was obtained by Rogelio Bernal using the double FSQ 10-cm equipment shown in Fig. 1 (right panel).}}
  \label{lugaro:fig1}
\end{center}
\end{figure}

 Amateur telescopes are also excellent instruments for obtaining deep narrow-band , wide- field images of nearby galaxies that can capture very fuzzy emission-line structures in their halos. This H$\alpha$-filter data  can even be gathered during full moon periods and from moderately light-polluted locations.  The recent detection of an ionized gas cloud in the outskirts of the famous M51 galaxy by Watkins et al. (2018) has inspired a systematic exploration of the outskirts of nearby galaxies (including the Andromeda galaxy)  with H$\alpha$-filters (Fig.~4; Mart\'\i nez-Delgado et al. in prep.). These vast clouds of gas could be related to tidal stripping of the ISM of the galaxy or starburst AGN winds (see C. Mihos´ review in this volume).
 

\section{Extra-galactic Stellar Tidal Streams}
\vspace{0.2cm}

Numerical cosmological simulations predict the stellar halos of massive galaxies contain an assortment of tidal phenomena exhibiting striking morphological characteristics, which should still be detectable  by deep imaging their outskirts. The most spectacular examples are long, tidal streams that wrap around the host galaxy and roughly trace the orbit of the progenitor satellite.

Extra-galactic stellar tidal streams cannot be resolved into stars with modest telescopes and thus appear as elongated diffuse light features that extend over several arc minutes as projected  on the sky. Their typical surface brightness is $\sim$ 27 mag/arcsec$^{2}$ or fainter, depending on the luminosity of the progenitor and the accretion time.  Detecting these faint structures requires very dark sky conditions and wide-field, deep images taken with exquisite flat-field quality over a wide area around the targets.

\subsection{The Stellar Tidal Stream Survey}

Over the last decade, the Stellar Tidal Stream Survey (STSS; Mart\'\i nez-Delgado et al. 2010; Mart\'\i nez-Delgado 2019) has obtained deep wide-field images of nearby spiral galaxies that previously showed signs of diffuse-light over-densities in shallower images from wide-area surveys (e.g. POSS-II; SDSS). This observational effort has yielded an unprecedented sample of bright tidal streams around spiral galaxies in the local Universe, with a total of $\sim50$  streams discovered so far\footnote{When the pilot phase of the STSS started in 2006, only 2-3 stream candidates were known, mainly from the pioneering work by David Malin using enhanced photographic plates (Malin \& Hadley 1997)}. The most conspicuous examples are shown in Fig.~5. This extraordinary variety of morphological specimens is compelling evidence in support of the hierarchical nature of structure formation predicted by cosmological models $\Lambda$CDM and mainly related to the stochastic nature of the satellite accretion process (Cooper et al. 2010). For example, in addition to {\it great-circle} features similar to the Sagittarius stream surrounding the Milky Way, the amateur data has revealed enormous structures resembling open umbrellas, with long, narrow shafts that terminate in a giant shell of debris extending several kiloparsecs into the halos, often on both sides of the host galaxy. The stream survey has also found isolated shells, giant clouds of debris floating within halos, jet-like ({\it spikes}) features emerging from galactic disks, giant plumes and large-scale diffuse structures that are possibly related to the remnants of ancient, already thoroughly disrupted satellites. Together with these remains of possibly long-defunct companions, these data has also captured surviving satellites in the act of tidal disruption (e.g. NGC 253; Romanowsky et al.\ 2016), revealing long tails departing from the progenitor.
 
\begin{figure}[t]
\begin{center}
 \includegraphics[width=0.85\textwidth]{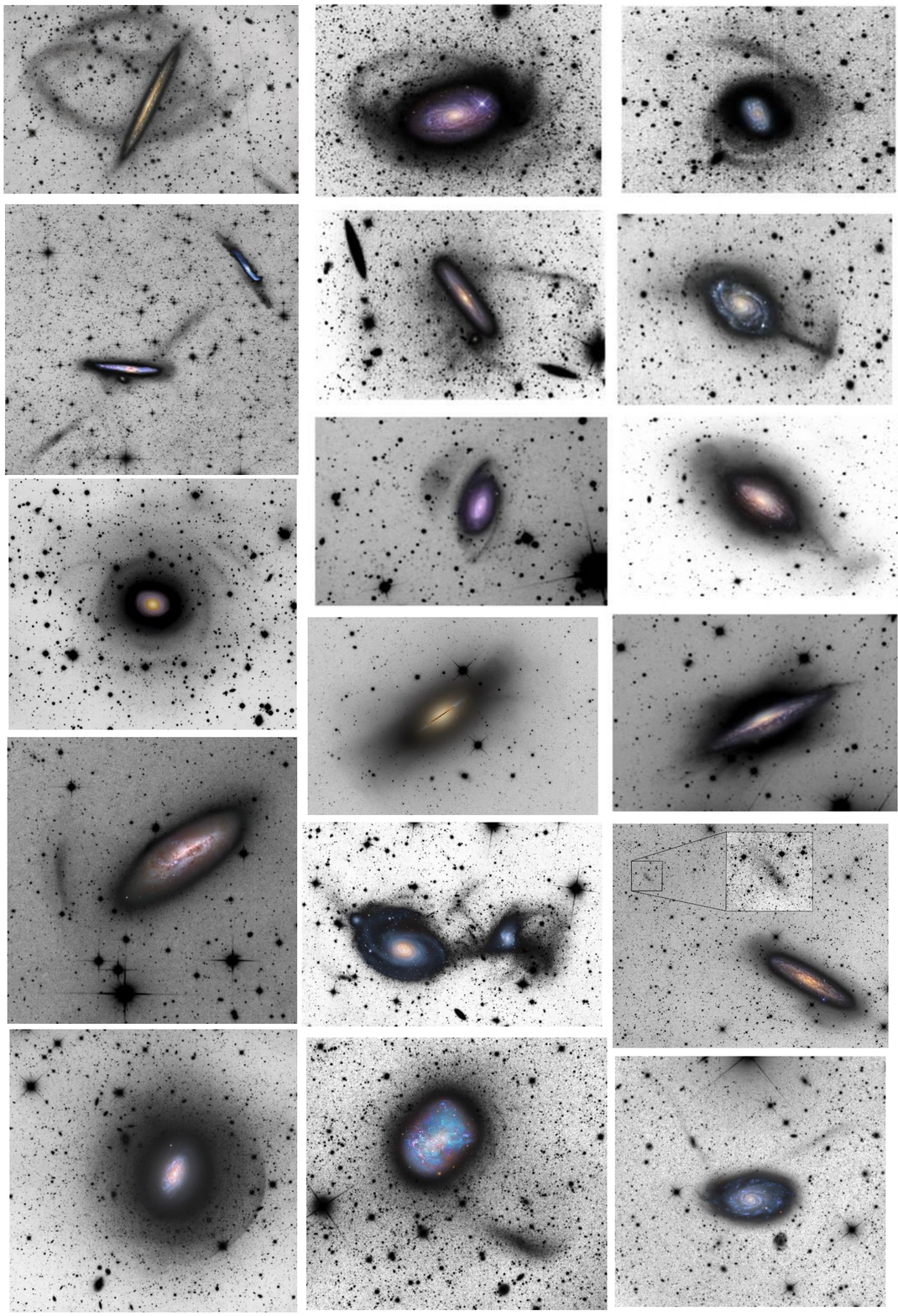} 
 \caption{Luminance filter images of nearby galaxies from the Stellar Tidal Stream Survey (2008-2018) showing large, diffuse light structures in their outskirts. A color inset of the disk of each galaxy has been over-plotted for reference. The typical SB of this structure is as faint as 26 mag/arcsec$^2$.}
  \label{lugaro:fig1}
\end{center}
\end{figure}

 The discovery of bright streams in some well-known, {\it classical} galaxies has opened a pathfinder for follow-up of some of the most striking tidal structures with much larger ground- and space-based telescopes. The results of this survey has also motivated detailed studies with intermediate-size telescopes of the most relevant structures discovered so far (NGC 4013: Mart\'\i nez-Delgado et al.\ 2009; M63: Chonis et al.\ 2011: NGC 5907: Laine et al.\ 2016) including N-body models to interpret their structured projected onto the sky (NGC 7600: Cooper et al.\ 2011; NGC 4651 : Foster et al.\ 2014;  NGC 4631: Mart\' \i nez-Delgado et al.\ 2015 ; NGC 1097: Amorisco et al.\ 2015; see Fig.~6).
 
 \begin{figure}
\begin{center}
 \includegraphics[width=1.0\textwidth]{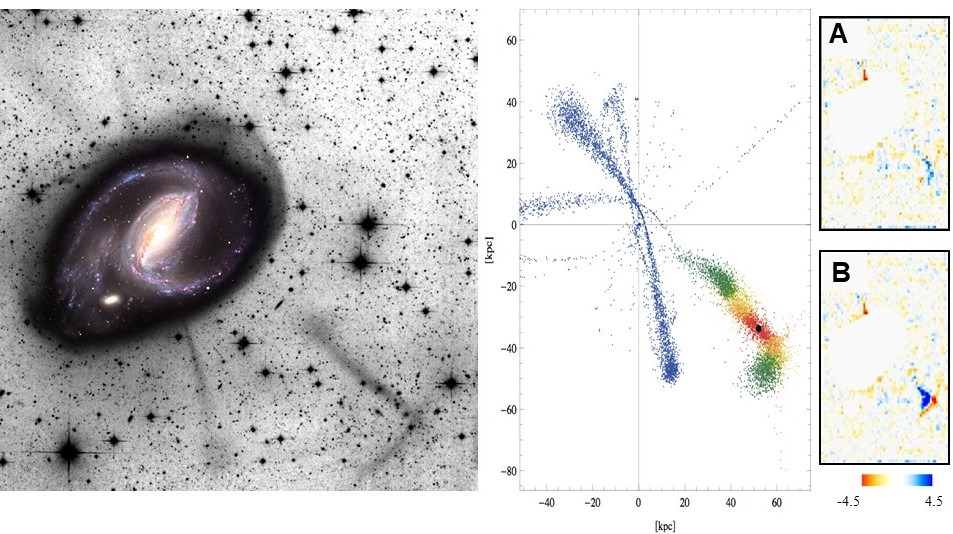} 
 \caption{\footnotesize{A spray-particle model fitting to the CHART32 0.5-m telescope image of the NGC 1097 tidal stream ({\it left panel}. The stellar material shedded by the progenitor is modeled using a purposely tailored modification of the {\it particle-spray} method (Gibbons et al.\ 2014). This technique faithfully reproduces the debris of an $N$-body disruption by ejecting particles from the Lagrange points of the progenitor, allowing to model the disruption in a few CPU-seconds. The best-fitting model ({\it center panel}) quantitatively reproduces the remnant location and the X-shape of the four `plumes' (Amorisco et al.\ 2015). The normalized residuals of that model to the surface brightness data ({\it right panels}) shows that the peculiar perpendicular (`dog-leg')} stream morphology can only be reproduced if rotation (case A) of the dwarf progenitor is included in the simulation.}
  \label{lugaro:fig1}
\end{center}
\end{figure}

An important highlight of this survey was the discovery of a stellar stream in the halo of NGC 4449,  an isolated Large Magellanic Cloud analog galaxy (Mart\' \i nez-Delgado et al. 2012; Rich et al. 2012). This is, so far, the lowest-mass primary galaxy with a verified stellar stream. This suggests that satellite accretion can also play a significant role in building up stellar halos around low-mass galaxies, and possibly in triggering starbursts. The NGC 4449 stream was resolved into a `river' of individual red giant branch stars with the 8-meter Subaru telescope, providing a tip of Red Giant Branch (TRGB) distance that confirmed its physical association with NGC 4449.  Another interesting result was the detection of massive star cluster candidates (similar to Omega Centauri in the Milky Way) embedded in the stellar stream of NGC 3628 (Jennings et al.\ 2015, see Fig.~7) and NGC 7242 (Mart\' \i nez-Delgado et al. in preparation), and the first evidence of enhanced star formation in the progenitor of an extra-galactic stream (NGC 5387: Beaton et al.\ 2014). 

 \begin{figure}[t]
\begin{center}
 \includegraphics[width=1.0\textwidth]{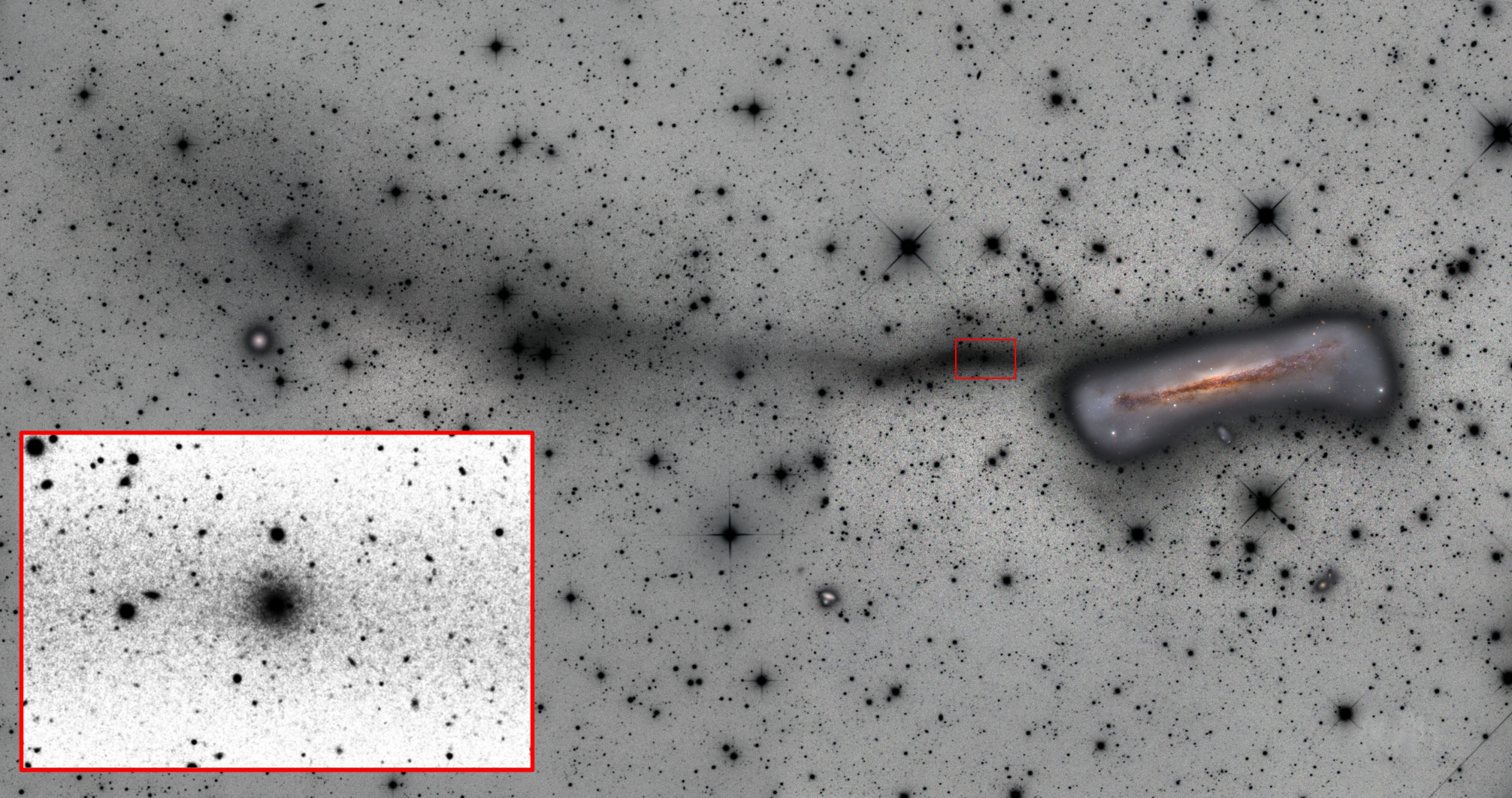} 
 \caption{\footnotesize{ An amateur mosaic image of the spectacular tidal stream of NGC 3628, one of the brightest streams in the local volume, obtained with a RC 0.5-m telescope by Mark Hanson. The inserted zoomed panel shows a Subaru 8.2-m telescope image of the cluster NGC3628-UCD1, a possible  Omega Cen-like analog embedded in this stream (Jennings et al.\ 2015). UCD1 is an example of a compact star cluster possibly forming from the nucleus of an accreted dwarf galaxy, suggesting that at least some of the massive globular clusters in the Milky Way and other galaxies  may be of tidal stripping origin.}}
  \label{lugaro:fig1}
\end{center}
\end{figure}

Galactic cirrus is the main limitation to undertake very deep probes of streams in stellar halos by preventing the attainment of surface brightness limits achieved with resolved stellar population studies of stellar halos of a few galaxies (e.g, M31: McConnachie et al. 2009; NGC 253: Greggio et al. 2014; NGC 5128: Crnojevic et al. 2016). To attain the surface brightness limit of the amateur images mentioned above (reached with a moderate exposure times of 6--10 hours using a luminance filter; see Sec.~2) and for the Galactic latitude range of the targets, significant contamination of extended filamentous features from Milky Way dust in a significant fraction of the data is expected, i.e. the typical images are similar to those shown in Fig.~5. However, longer exposure times ($>$ 20-30 hours) reveal that the complex field of very faint cirrus (see Fig.~8) is ubiquitous, even visible at high Galactic latitudes (e.g. the Virgo cluster). These foreground dust structures are visible at extremely low surface brightness regime ($>$ 29.5 magn/arcsec$^{2}$) and prevent the detection of most stellar halo sub-structures (with surface brightness fainter than 30.5 magn/arcsec$^{2}$)  predicted by cosmological models (e.g. see left panel in Fig.\ 9). This also means that, in the majority of cases, it is only possible to detect one (that corresponds to the brightest structure that inhabits in its halo) or no streams in each galaxy observed with amateur telescopes. A comprehensive discussion about how to disentangle (and in some cases decontaminate) the presence of cirrus in the target fields is given in the C. Mihos´ review (this volume). In addition, color information of these features can be used to find out their origin, following the new approach given by Roman et al. (2019).

 \begin{figure}
\begin{center}
 \includegraphics[width=1.0\textwidth]{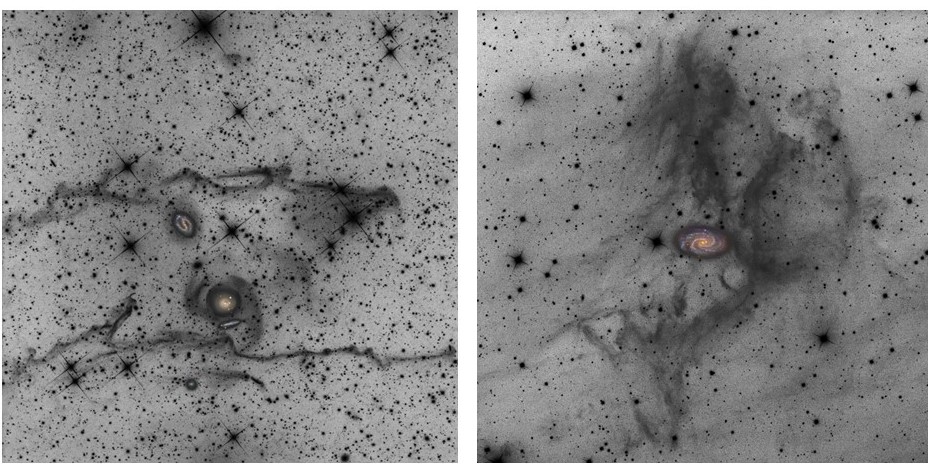} 
 \caption{\footnotesize{ Deep images of two nearby galaxies located in a stellar field seriously contaminated by Galactic cirrus. ({\it Left panel}): A very narrow, twisted and filamentous cirri around NGC 2634, a post-merger elliptical galaxy. {\it Right panel}): A very complex structure of cirrus surrounding the nearby spiral galaxy NGC 918. Both images were taken with a RC 0.5-m telescope by Mark Hanson, with exposure times of 10 hours and 5 hours respectively. }}
  \label{lugaro:fig1}
\end{center}
\end{figure}

The final data produced by the STSS during the next years will help to complete the census of stellar streams in the Local Volume, and so to understand if the frequency and properties (e.g.\ surface brightness, morphology, progenitor) of tidal streams around nearby galaxies are  consistent with predictions from $\Lambda$CDM simulations.

\begin{figure}[t]
\begin{center}
 \includegraphics[width=1.0\textwidth]{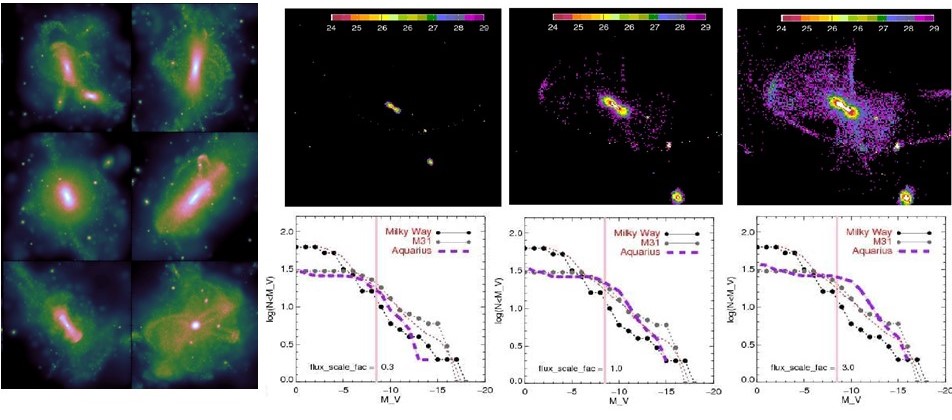}

 \caption{\footnotesize{({\it Left panels}): Expected halo streams around a Milky Way - like galaxy from the Coco cosmological simulations (created with the technique presented in Cooper et al. 2013, 2017). The panels show an external perspective of several simulated galaxies with streams within the hierarchical framework. They illustrate a variety of typical accretion histories for Milky Way-type galaxies. Each panel is 300~kpc on a side and show a stellar mass density range of $10^{1}$ to 10$^{8}$ ${\rm M}_{\odot}/kpc^{2}$. ({\it Right panels}): A representative sample of tidal streams in a large number of galaxies can constrain models of galaxy formation. {\it Bottom row}: The luminosity function (LF) of satellites in a simulated Milky Way analogue (Aquarius-A; Lowing et al. 2015) compared to observational data from the Milky Way and M31. From left to right we simply rescale the luminosity of all  satellites in the model by a factor of 0.1, 1 and 3.0 respectively. Given the limited data available, any of these predictions could be judged an acceptable match. {\it Top panels} show how this rescaling affects the surface brightness distribution of the host galaxy's stellar halo.}}
  \label{lugaro:fig1}
\end{center}
\end{figure}


\subsection{Comparison with cosmological models}

The census of stellar streams in nearby spiral galaxies will provide a comprehensive statistical foundation necessary to directly compare observed stellar halo substructure with  high-resolution cosmological simulation predictions. 
A complete theoretical understanding of complex nonlinear features like stellar tidal streams and their contribution to the stellar halos in nearby galaxies requires state-of-the-art cosmological simulations. In this context, the {\textbf{Copernicus Complexio (\textit {CoCo})}} cosmological $N$-body simulations from a recently completed project of the VIRGO consortium (Hellwing et al. 2016) provides both high mass resolution and a representative analogue of the Local Volume. The \textit{CoCo} simulations (Fig.~9, left panel) have sufficient volume (a spherical region of $\sim25$~Mpc embedded in a lower-resolution box of $100$~Mpc/side), resolution (a particle mass of $1.6\times10^{5} M_{\odot}$) and dynamic range (a softening scale of $330$pc, comparable to the width of the Milky Way's Orphan stream) to draw representative samples of local galaxies that focus on the tidal debris from dwarf galaxies comparable to the classical satellites of the Milky Way. 

Using particle tagging methods (e.g. Cooper et al. 2010, 2013, 2017), cosmological $N$-body models can be used to study the statistics of low surface brightness structures at a resolution comparable to or better than state-of-the art hydrodynamical simulations, but at only a fraction of the computational cost. For example, the \texttt{Galform} semi-analytic
model of galaxy formation (Lacey et al.\ 2016) has been used to predict the evolution of stellar mass, size, and chemical abundances in every \textit{CoCo} dark matter halo, constrained by comparisons to the large-scale statistics of the cosmological 
galaxy population (for example, optical and infrared luminosity functions). Individual dark matter particles in \textit{CoCo} can then be “painted” with single-age stellar populations according to a dynamical prescription that specifies the binding energy distribution of stars at the time of their formation (see Cooper et al. 2017 for details of the technique). With this approach, the individual star formation history of each satellite (and hence properties such as stellar mass, luminosity and metallicity) can be studied alongside the full phase-space evolution of its stars. Crucially, for comparison with the observations, the simulation data can be projected and realistic observational effects included (e.g. sky noise, flat-field corrections, surface brightness limit, etc.) to generate mock images that we can analyze with the same techniques applied to the real data.

Fig.~9 (right panel) shows how a representative sample of tidal streams in a large number of galaxies can constrain models of galaxy formation. The bottom panels show the luminosity
function (LF) of satellites (dashed line) in a simulated Milky Way analogue galaxy  from the Aquarius-A simulations (Lowing et al. 2015) compared to observational census of companions around the Milky Way (black line) and M31 (gray line). Predictions on this scale (for both hydrodynamical and semi-analytic models) are sensitive to modeling choices that are not well constrained by calibration to data on larger scales.  In this model, the satellite LF shape and amplitude are primary determined by a combination of five parameters (three describing supernova-driven feedback and two describing the effects of cosmic re-ionization).  From left to right bottom panels,  the luminosity of all satellites is re-scaled in the model by a factor of 0.1, 1 and 3.0 respectively, to approximate the effects of varying relevant interdependent physical parameters (see Font et al. 2011). Given the limited data available for the Local Group, any of these predictions could be judged as an acceptable match to the data. The top panels shows three snapshots for the same model and surface brightness cut-off similar to that of amateur data obtained in the STSS ($\sim$ 29 magn/arcsec$^{2}$; see Sec.~3.1). This illustrates how the re-scaling in the LF clearly affects the detectability of stellar streams and other low surface brightness features in the host galaxy's stellar halo. This means that large variations in the faint structure observed above the surface brightness limit of our survey follow from changes in the total abundance of bright satellites accreted by the system over cosmic time. New generations of cosmological simulations are specifically targeting this faint
regime because of its importance to tests of $\Lambda$-CDM. Thus, a systematic survey of stellar streams in the local universe will offer new constraints on such models that are highly complementary to those obtained from satellite LFs in nearby galaxies (and with very different observational systematic errors).



\begin{figure}[t]
\begin{center}
 \includegraphics[width=1.0\textwidth]{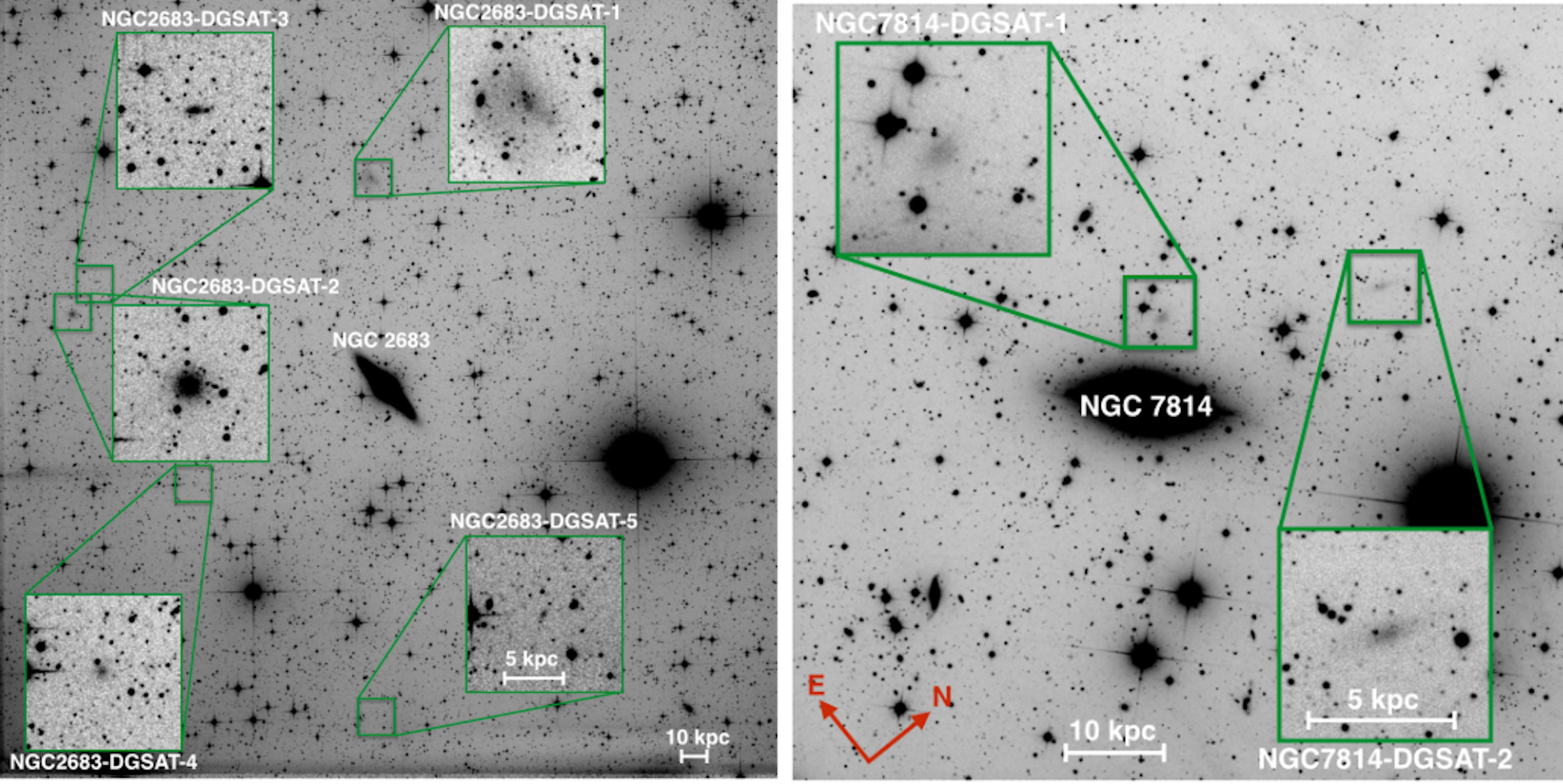} 
 \caption{\footnotesize{Dwarf galaxy satellites discovered in the nearby Milky Way-like galaxies NGC 2683 and NGC 7814 as part of the DGSAT survey. These images were taken by Karel Teuwen with the ROSA 0.40-meter telescope in the Southern Alps (Javanmardi et al. 2015; Henkel et al. 2017).}}
  \label{lugaro:fig1}
\end{center}
\end{figure}

\section{Completing the Census of Faint Dwarf Satellites around Nearby Galaxies}


Searches for stellar streams with deep amateur imaging have also lead to the discovery of numerous faint dwarf satellites around a handful of nearby spiral galaxies. New satellites have been found in almost all the galaxies explored so far, suggesting the previous images of these systems situated within 20 Mpc were too shallow for undertaking a complete census of their dwarf satellite population.

There are a few groups of organized astrophotographers devoted to the search of these low-surface brightness galaxies with amateur telescopes. The {\it Dwarf Galaxy Survey with Amateur Telescopes} (DGSAT; Javanmardi et al. 2016; Henkel et al. 2017) is in fact a "spin-off" of the STTS, since it uses their same telescopes and data (Fig.~10). An observing project with a similar goal is the {\it Tief Belichtete Galaxies} (TBG)) by a group of German and Austrian astrophotographers  (Karatchensev et al. 2015). Another  project (but mainly composed by professional astronomers) is the {\it Halos and Environments of Nearby Galaxies} (HERON) (Rich et al. 2019), that only uses a dedicated  0.7-m amateur telescope situated near Frazier Park (California). Their survey comprised 119 galaxies from the 2MASS nearby bright galaxy catalogue, but the list of all the discovered companions around them has not been published yet.

The dwarf candidates can be detected semi-automatically with a pipeline designed to calibrate the luminance images taken by these amateur telescopes, search for dwarf galaxy candidates, and extract their observed photometric and structural parameters. All  of  the dwarf satellite candidates detected in these surveys have  very  low  surface  brightness  ($>$25  mag/arcsec$^{2}$),  and thus cannot be detected  in the shallow images  from  large-scale  surveys  like  the  SDSS or  PanSTARRs\footnote{New generation imaging surveys, like the Legacy surveys or Dark Energy Survey, provides deep data comparable to that from amateur telescopes for this dwarf galaxy hunting.}.  Their  surface  brightness  are in the range 25.3--28.8 mag/arcsec$^{2}$ and their  Sersic  indices  are n$<$1, which  are  in the range of  those found  in the  dwarf companions in the spirals of the  Local  Group  (McConnachie  2012). 

However, further observations are required to confirm the discovered galaxies are dwarf satellites of their nearby massive galaxies. Their surface brightness   makes  it  very  difficult  to  undertake follow-up observations for obtaining their radial velocities even  for 8-meter  class  telescopes.  
Assuming  that  they  are  dwarf  satellites  of  their neighbor (in  projection)  massive  galaxies,  their $r$-band  absolute  magnitude and effective radius is −15.6$<M_{r}<$ −7.8 and 160 pc $<R_{e}<$ 4.1 kpc (Javanmardi et al. 2016), respectively. This would mean that amateur telescopes are able to  detect systems with similar observed properties to those of the “classical” dwarf spheroidal galaxies around the Milky Way.
This also showed the potential of small-sized telescopes as successful tools for probing these low surface brightness systems in the nearby Universe.

\begin{figure}[t]
\begin{center}
 \includegraphics[width=1.0\textwidth]{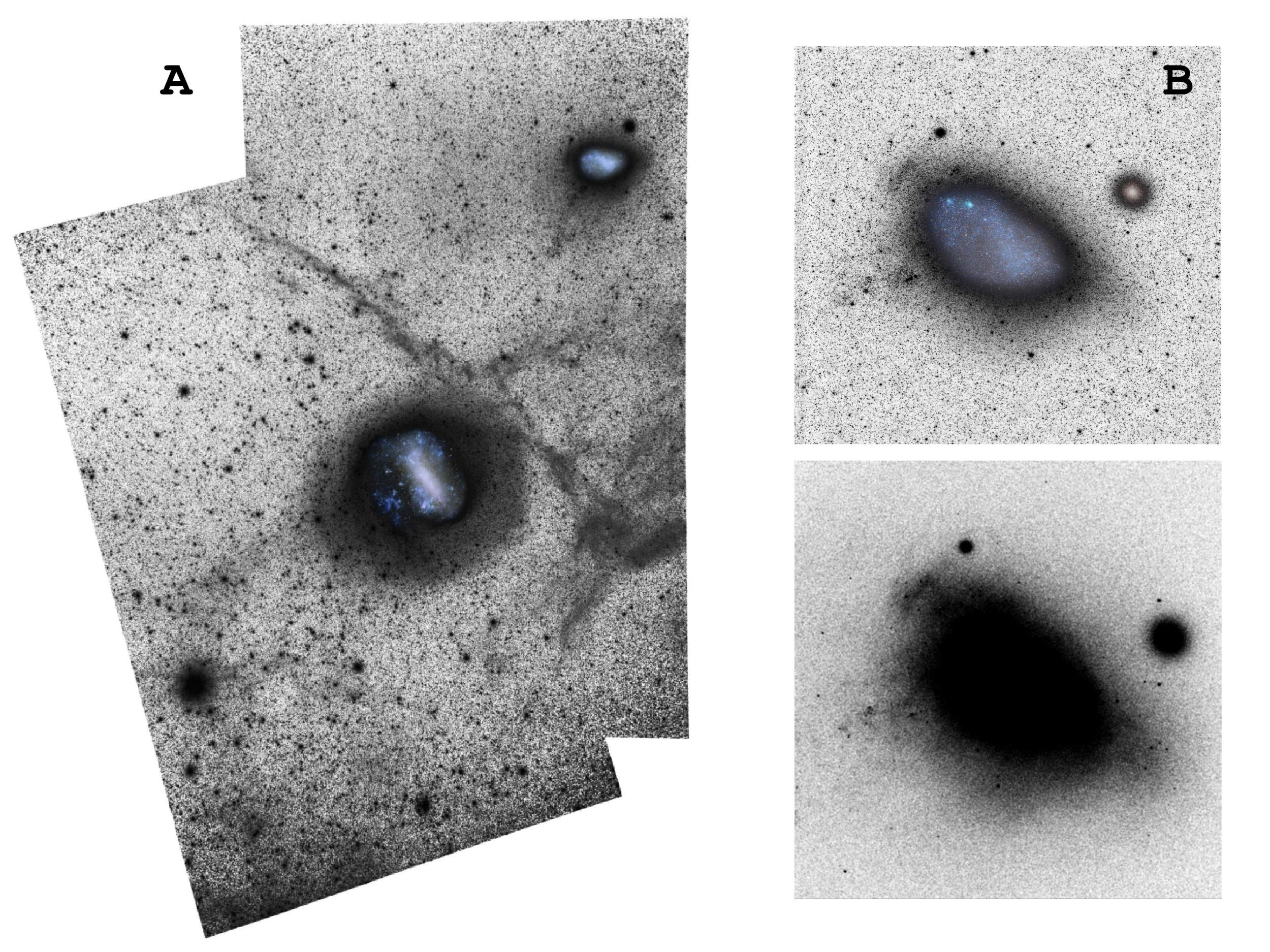} 
 \caption{\footnotesize{{\it Left panel}: Full wide-field mosaic of the Magellanic Clouds obtained with a  Canon EF 50mm f/1.4 USM prime lens as described in Sec.~5 from La Silla Observatory (Chile) by Yuri Beletsky. {\it Right top panel}: Image of the SMC obtained with a Canon EF 200 mm f/2.8L lens attached to a Canon E0S6D (ISO1600) camera  from Hacienda Las Condes (Chile). Image taken by Bernhard Hubl. The total field of view is 10.2º $\times$ 6.8º, with a pixel scale of
6.7 arcsec~pixel$^{-1}$. A shell-like over-density (composed by young stars) is clearly visible in the North-East side of the Cloud. ({\it Right bottom panel}: For a comparison, a stellar density map of the SMC obtained  with all the {\it Gaia} DR2 (Gaia Collaboration et al. 2018) sources. The mentioned shell and other fainter features (see Mart\'\i nez-Delgado et al. 2019) are clearly visible in the $Gaia$ data.}}
  \label{lugaro:fig1}
\end{center}
\end{figure}

\section{Ultra-Deep Imaging of the Magellanic Clouds: Imprints of Tidal Interactions between the Clouds}

One important approach toward understanding the formation and evolutionary history of the Magellanic Clouds is through wide-field, deep mapping of
the Clouds' outskirts. These regions can still contain clues about the times and nature of past interactions between the Clouds that could have left
visible imprints in the stellar component such as distortions, clumps, arcs and related over-densities  (Pieres et al.\ 2017; Mackey et al.\ 2018). With that purpose, Mart\'\i nez-Delgado et al. (2019) carried out a deep, wide-field imaging survey of stellar substructures in the periphery of the Magellanic Clouds,  inspired by the photographic plate work by de Vaucouleurs dating back to the fifties (de Vaucouleurs \& Freeman 1972; see Fig. 1 in Mart\'\i nez-Delgado et al. 2019). This project used low-cost telephoto lenses with the goal of obtaining unresolved images of the Clouds, using the observational approach described in Sec. 2.

The panoramic view of the LMC and SMC fields shown in Fig.~11 was obtained during two observing runs in August and September 2009 at
ESO La Silla observatory. This mosaic of two images was done using a portable setup consisting of a SBIG STL-11000M CCD camera and a Canon EF 50mm
f/1.4 USM prime lens, which yielded a  of 39 x 27 deg and resulting pixel scales of 37"/pixel. Each image set consists of deep
multiple exposures obtained through a Baader Luminance filter (4000 \AA $< \lambda <$7500 \AA) with a total exposure time of 290 minutes. 

A conspicuous tail of young start is visible emanating to the East of the SMC, extending ~ 6 degrees toward the LMC and overlapping the Magellanic Bridge. However, this known structure was most likely formed in situ rather than being tidally stripped. The LMC also shows an asymmetric structure, 
revealing the existence of stellar arcs and multiple spiral arms in the northern side, with no comparable counterparts in the South.  Besla et al. (2016) 
compared this image to theoretical simulations of the LMC disk perturbations resulting from its past interactions  with the SMC. They found the origin of these northern structures is a repeated dwarf-dwarf interaction with the SMC. The rest of coherent filaments visible in the field (and first reported in the photographic plates study by de Vaucouleurs \& Freeman 1972) are likely associated with extended structures of Galactic cirrus (see Sec.~2), that are abundant in deep imaging even at this high  galactic latitude (see discussion in Sec.2). 

 A second striking feature barely visible in these low-resolution images is a coherent shell-like over-density feature located at $\sim$ 1.9º from the center of the SMC in the north-east direction. To ensure this shell-like feature was not an artifact or a 
reflection\footnote{This feature was also previously detected in photographic studies of the Clouds 
dating back to the  1950s (e.g. see Figure 12b in de Vaucouleurs \& Freeman 1972).}, a confirmation wide-field image of the SMC was obtained with a different Canon EF 200 mm f/2.8L lens attached to a Canon E0S6D (ISO1600) camera
on 2015 October 10 from Hacienda Los Andes (Chile). 

Fig.~11 (right top panel) shows the resulting image, with a total exposure time of 156 minutes, compared with 
a stellar density map of the SMC made with all the {\it Gaia} DR2 (Gaia Collaboration et al. 2018) sources (right bottom panel). The spatial distribution of the resolved stellar populations (Mart\'\i nez-Delgado et al. 2019) of this elongated
over-density shows that it is actually a brighter optical part of a more extended, spiral-arm-like structure mainly traced by blue, young stars of the Cloud. The over-density also contains at least nine young star clusters with ages tightly clustered at 175 Myr. However, there is no counterpart of this over-density in the distribution of the older stars in this area of the SMC. This suggests the
nature of this over-density seems to be different from the one discovered at  $\sim$ 8 degrees north of the SMC by Pieres et al.\ (2017)
 (mainly composed of intermediate-age stars) and from the morphologically similar (but older and larger) substructure described by Mackey et al. (2016).
This contrast between the small-scale spatial structures in the young populations and the smooth distribution of the intermediate-age and old populations strengthens the argument that this feature was not produced via tidal effects.

In fact, the analysis of the stellar content of the shell using SMASH photometry indicates the main difference between the stellar population of the shell and the underlying SMC population is a recent period of enhanced star formation during the last $\simeq$ 1 Gyr, likely peaking at $\sim$ 150 Myr, as indicated by the clusters
 and Cepheids age distribution (Mart\'\i nez-Delgado et al. 2019). Thus, the young age of this SMC could be related to the possible recent interaction with the LMC about 100 to 300 Myr ago, which left its signature in the
young stellar populations in the Magellanic Bridge (e.g., Skowron et al.\ 2014) and in the age distribution of young star clusters
in both Clouds (e.g., Glatt et al.\ 2010). This is also consistent with the recent proper motion study by Zivick et al. (2018), that found the LMC and SMC
have had a head-on direct collision 147 $\pm$ 33 Myr ago, and in very good agreement with the peak time of star formation enhancement found in the SMC shell.

\section{ Searching for quenched faint dwarf galaxies around the Local Group}

The $\Lambda$-CDM paradigm predicts a large number of small dark matter halos inside and around the Local Group, but it is unclear how many of them  are associated with luminous baryons in the form of stars. It is thus of broad interest for galaxy formation theory to carry out a full inventory  of the numbers and properties of dwarf galaxies, both satellites and isolated ones.
The study of quenched isolated dwarfs composed exclusively by old stars outside the Local Group is fundamental to investigate the effect of their environments versus internal processes on their star formation histories. Geha et al.(2012) found quenched dwarf galaxies  are always located within 1.5~Mpc of a more massive host.

A systematic search for these extremely faint systems around the Local Group using  short-focal ratio telescopes and telephoto lenses with the observational approach described in Sec.~2 would help constrain their actual number at lower surface brightness regimes and test modern cosmological scenarios that assume a different origin for this kind of galaxies. Because of the independence of the surface brightness on the distance in the nearby  universe, this equipment provides a fast way to explore vast sky areas and look for extremely faint dwarf galaxies beyond the Local Group, up to distances of 10-15 Mpc (e.g. Danieli, van Dokkum \& Conroy 2018), that could still remain undetected in large scale optical and radio surveys due to their extremely faint surface brightness and low gas content.

\begin{figure}[t]
\begin{center}
 \includegraphics[width=1.0\textwidth]{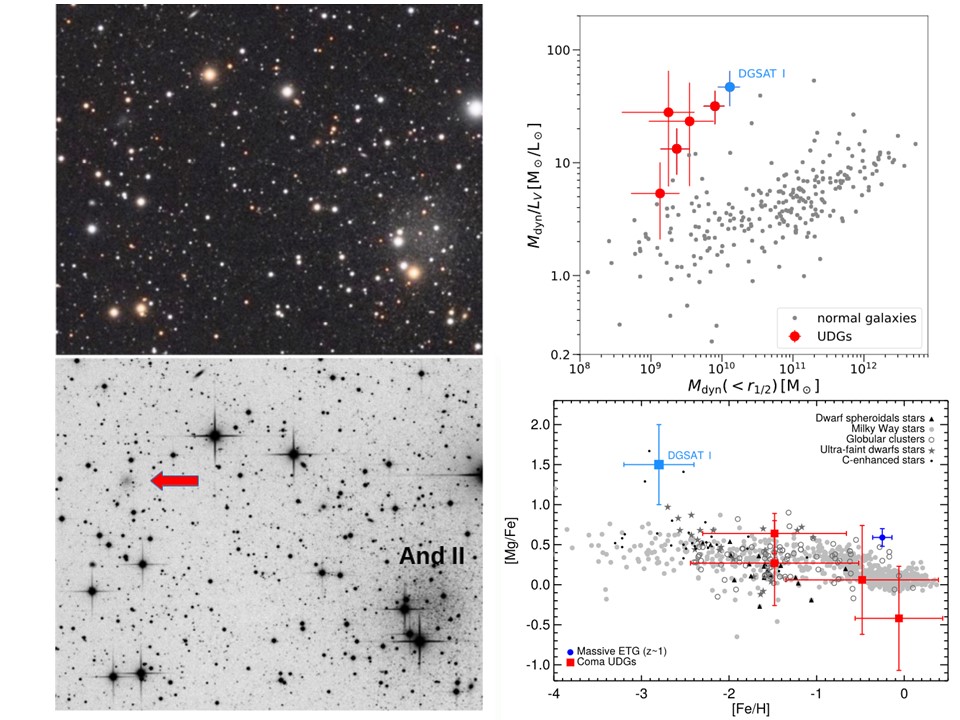} 
 \caption{\footnotesize{({\it Left panel)} Confirmation image of the ultra-difuse galaxy DGSAT~I taken by Jordi Gallego with a 106-cm apocromatic refractor from the Parc Astron\'omic del Montsec. The galaxy was discovered in the proximity of the M31 dwarf satellite And II; ({\it Right panel}): A follow-up study of DGSAT~I  by  Martin-Navarro et al. 2019 ({\it top}) confirmed the properties of this galaxy are challenging to interpret within the galaxy formation theory. {\it Upper panel}: the dynamical mass-to-light ratio vs. the dynamical mass for DGSAT~I  and  other  UDGs  (red  points), showing a large excess  of  dark matter with respect to normal galaxies. {\it Bottom panel}: the [Mg/Fe] ratio vs. iron metallicity of DGSAT~I compared with those for field stars, Milky Way globular clusters and other systems. DGSAT I hosts extremely magnesium-enhanced [Mg/Fe] abundance ratio, with lower matallicity and higher [Mg/Fe].}}
  \label{lugaro:fig1}
\end{center}
\end{figure}

\subsection{The discovery of DGSAT~I, an isolated field ultra-diffuse galaxy}

The most important discovery so far of the DGSAT project (see Sec.~4) is a field ultra-diffuse galaxy named DGSAT I (Mart\'\i nez-Delgado et al. 2016).  Its unresolved appearance in the amateur (Fig~12, left panel) and Subaru follow-up images,  structural  properties  and  absence of emission lines initially suggested an interesting case of an isolated dwarf galaxy $\sim$ 10 Mpc beyond the Local Group with  a  surface  brightness  and  structural  properties  similar  to  those  of  the  classical  Milky Way  dSph companions.  However, spectroscopic follow-up observations with the BTA 6-meter telescope yielded a radial velocity of 5450$\pm$40 km s$^−1$. This revealed DGSAT I is actually a background system placed at a distance of $\sim$ 78 Mpc and likely associated with an outer filament of the Pisces-Perseus super-cluster  projected in this direction of the sky. Unlike other UDGs found in the Coma, Virgo and Fornax galaxy clusters (see E.K. Grebel review, this volume) at that time,  DGSAT  I is a rare example of isolated UDG that resides in a low density environment. 

The isolation of DGSAT~I offers an opportunity to study the secular evolution of this type  of  object and to explore different  formation scenarios for UDGs without confounding environmental effects. Since its discovery, DGSAT~I has been observed with the Hubble Space Telescope (Romanowsky et al. in preparation), Keck Cosmic Web Imager (Martin-Navarro et al. 2019), Spitzer Space Telescope (Pandya et al. 2018) and it has also inspired theoretical simulations with the goal of explaining its formation (Di Cintio et al. 2017). Its estimated stellar mass is 4.8$\times$ 10$^8$ M$_{o}$, and it shows little sign of cold gas or current star formation (Papastergis et al. 2017). Its  measured velocity dispersion of $\sigma$ $\sim$56 km s$^−1$ indicates a high dark matter-to-stellar mass ratio. Moreover, DGSAT I also exhibits an extended star formation history that would be in contradiction with its extremely magnesium-enhanced stellar populations (about ten times higher than the most magnesium-enhanced stellar systems discovered to date; Martin-Navarro et al. 2019; see Fig~12 right top panel). The  exact  mechanism(s)  regulating  the chemical evolution of DGSAT I  is still unknown, but the lack of iron suggests an undeveloped chemical evolution that is typical of the star formation conditions in primeval galaxies.  Thus, follow-up detailed studies of its stellar population may provide a window into the early Universe.

\begin{figure}[t]
\begin{center}
 \includegraphics[width=1.0\textwidth]{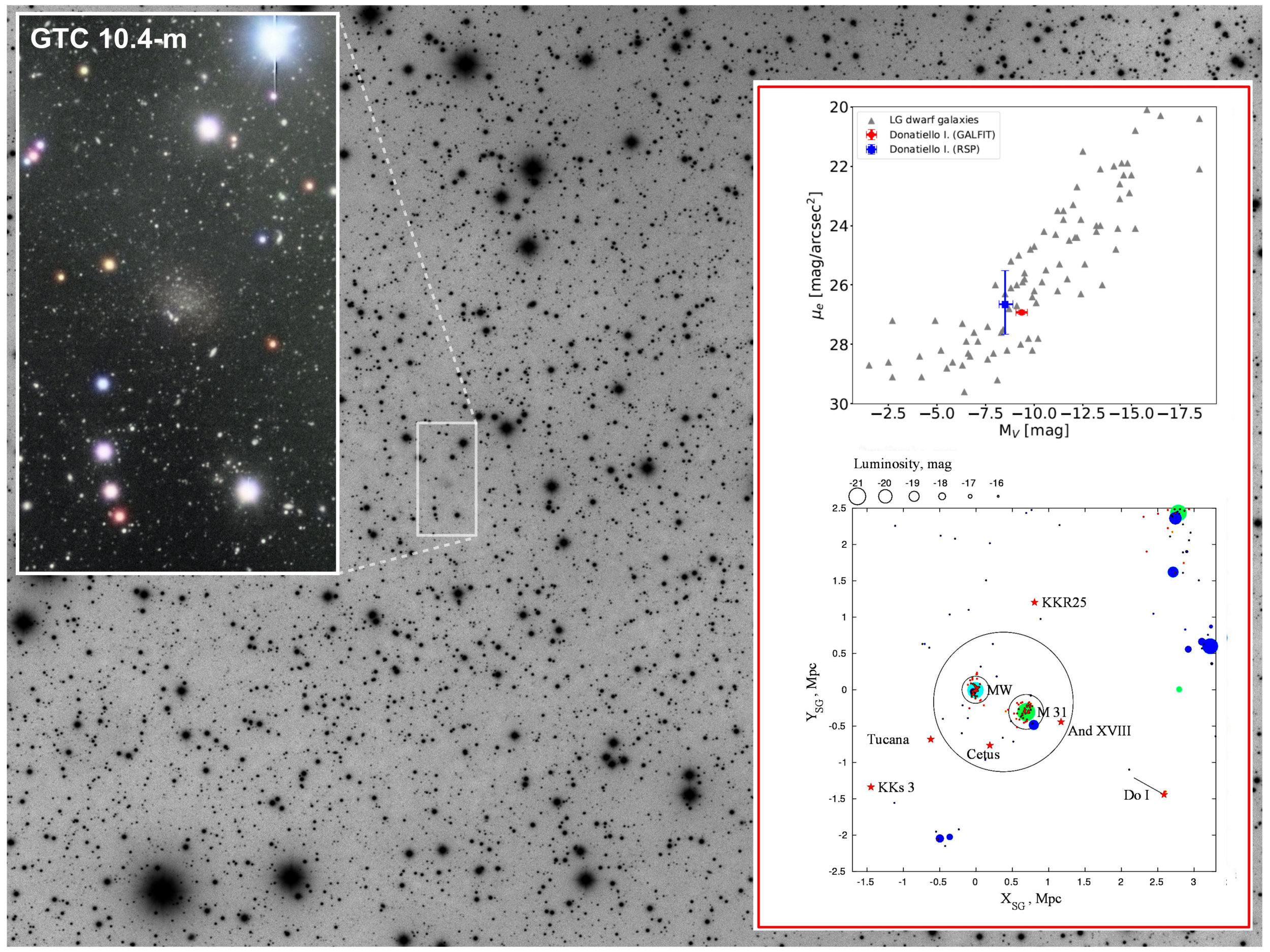} 
 \caption{\footnotesize{ ({\it Left panel}): Image of the dwarf galaxy Do~I taken with a Takahashi FSQ106  refractor at f/3.6 and a commercial CCD camera QHY90A at the Avalon Merlino Remote Observatory by Giuseppe Donatiello. The total exposure times were  24300 sec in the Luminance filter. The inner zoomed panel shows the  $r$-band image of Do~I obtained from Gran Telescopio de Canarias 10.4 meter telescope observations. {\it Top right panel}: A comparison between the $V$ band surface brightness in mag arcsec$^{-2}$ vs. absolute $V$ band magnitude.properties of Do~I and the known dwarf galaxies of the Local Group (McConnachie 2012). {\it Bottom right panel}: Super-galactic coordinate map of the distribution of galaxies within a 5~Mpc cube roughly centered on the Local Group. The size of filled circles is scaled to the luminosity of a galaxy.  The big circle encompassing the Local Group corresponds to the zero-velocity surface, which separates the collapsing region of space around the Local Group from the cosmological expansion. Do~I and other known isolated dSph systems are marked with a red star. The short line from Do~I marks its distance uncertainty from 2.5--3.5~Mpc. In this map, the position of the SO galaxy NGC~404 almost overlaps with those of Do~I at 3.5 Mpc. }}
  \label{lugaro:fig1}
\end{center}
\end{figure}

\subsection{Donatiello I: an isolated quenched dwarf galaxy behind Andromeda?}

The number of known very isolated dSph systems is very small, not exceeding 5 of more than 1000 of galaxies found in the Local Volume. However, the majority of them could still remain undetected in large scale optical and radio surveys due to their extremely faint surface brightness and low gas content.

A very interesting example in this context is Donatiello ~I (Do~I; Mart\'\i nez-Delgado et al. 2018), a low-surface brightness ($\mu_{V}=26.5$ mag arcsec$^{-2}$) stellar system located at a projected distance of one degree from Mirach ($\beta$ And) (Fig.~13, left panel) and discovered by the italian amateur astronomer Giuseppe Donatiello. This dwarf galaxy was first found during a deep image visual inspection of the Andromeda galaxy region taken with a 12.7-cm refractor ED127mm . 

This discovery was confirmed by a visual inspection of the SDSS DR9 images and follow-up observations using the Telescopio National Galileo 3.6-m and the Gran Telescopio de Canarias 10.4-m (La Palma, Spain; Fig.~13). The color--magnitude diagram revealed this system is beyond the Local Group with a stellar content similar to  the "classical" Milky Way companions Draco and Ursa Minor. The absence of young stars and HI emission in the ALFALFA survey are also typical of quenched dwarf galaxies. Crowding effects in the images did not enable the secure establishment of Do I's distance but it provided a tentative distance modulus for this galaxy of $(m-M)=27.6 \pm 0.2$ (3.3 Mpc) and an absolute magnitude of $M_{V} =-8.3$ (Fig. 13, right top panels). Its projected position and distance of Do~I are still consistent with being a dwarf satellite of the closest S0-type galaxy NGC 404
 ("Mirach's Ghost"). A more accurate distance (using {\it Hubble Space Telescope} imaging) and a radial velocity measurement are needed to shed additional light on this interaction scenario and definitively determine if Do~I is a NGC 404 companion. Alternatively, it could be one of the most isolated quenched dwarf galaxies reported so far (see Fig.~13 right bottom panel) behind the Andromeda galaxy . 

\section*{Acknowledgments}

I thank to the organizers of the IAU 355 Symposium for their kind invitation to give this review, to Giuseppe J. Donatiello for his help with the composition of some figures of this proceeding and to Andrew P. Cooper and R. Jay GaBany for comments. I also express my gratitude to the world-class astrophotographers who have contributed to this project during the last decade: R. Jay GaBany, Ken Crawford, Mark Hanson, Yuri Beletsky, Adam Block, Fabian Neyer, Rogelio Bernal, Jordi Gallego, Karel Teuwen, Johannes Schedler, Bernhard Hubl, Giuseppe J. Donatiello, S. Mazlin,  Michael Sidonio, Daniel Verschatse, Josep M. Drudis, Vicent Peris, Alvaro Ib\'a\~nez and Manuel Jim\'enez. I also thank to Andrea Pistocchini for his image of the Leo I dwarf, A. Romanowski for the Subaru image of NGC 3628 and Andrew P. Cooper for the  CoCo cosmological simulation particle-tagging snapshots. I acknowledge support by Sonderforschungsbereich (SFB) 881 ``The Milky Way System'' sub-project A2 of the German Research Foundation (DFG). I acknowledge support from the Spanish Ministry for Science, Innovation and Universities and FEDER funds through grant AYA2016-81065-C2-2.

\begin{discussion}
\discuss{J. S\'anchez Almeida}{Could your various setups also detect faint diffuse emission lines\,?}
\discuss{D. Mart\'{\i}nez-Delgado}{Yes, in fact we have started a systematic survey with a H$\alpha$ filter to search for ionized gas clouds around nearby spiral galaxies. The main limitation for using other narrow filters ([OIII], [SII]) is to find a manufacturer that provides the suitable size for the amateur equipment at a reasonable price.}

\discuss{A. Gil de Paz}{What can you say about the rotation of the progenitors as a function of the
dark matter slope in these systems\,?}
\discuss{D. Mart\'{\i}nez-Delgado}{We have only found indirect evidence of rotation in the progenitor of the NGC 1097 stream (see Fig. 6) so far. Thus,  it is too early to get any conclusion about this issue yet.}


\discuss{F. Buitrago}{Is it possible to estimate the age of tidal tails\,?}
\discuss{D. Mart\'{\i}nez-Delgado}{ I think you mean the accretion time of the progenitor of the tidal stream. All the observed structures around nearby galaxies are relatively young, formed in the last few Gigayears (like the Sagittarius tidal stream in the Milky Way). The accretion times are only possible to estimate in the most complex streams displaying multiple features (e.g. NGC 5907, NGC 1097, NGC 4651) by means of the comparation of their structure projected onto the sky with theoretical simulations.}

\end{discussion}

\end{document}